# REAL TIME ESTIMATION IN LOCAL POLYNOMIAL REGRESSION, WITH APPLICATION TO TREND-CYCLE ANALYSIS

By Tommaso Proietti and Alessandra Luati

*University of Rome "Tor Vergata" and University of Bologna*

The paper focuses on the adaptation of local polynomial filters at the end of the sample period. We show that for real time estimation of signals (i.e., exactly at the boundary of the time support) we cannot rely on the automatic adaptation of the local polynomial smoothers, since the direct real time filter turns out to be strongly localized, and thereby yields extremely volatile estimates. As an alternative, we evaluate a general family of asymmetric filters that minimizes the mean square revision error subject to polynomial reproduction constraints; in the case of the Henderson filter it nests the well-known Musgrave's surrogate filters. The class of filters depends on unknown features of the series such as the slope and the curvature of the underlying signal, which can be estimated from the data. Several empirical examples illustrate the effectiveness of our proposal.

**1. Introduction.** One of the key issues economists have faced in characterizing the dynamic behavior of macroeconomic variables, such as output and inflation, is separating the longer-term component from the transitory one. Key measurements such as dating the business cycle turning points and more generally the assessment of the underlying trend call for signal extraction methods that separate the two components. Many methodologies are available for the task, ranging from nonparametric methods based on the notion of a band-pass filter and on wavelet methods [Percival and Walden (2000)], kernel estimation and local polynomial modeling [see, e.g., Fan and Gijbels (1996)] semiparametric methods based on spline smoothing and mixed models [see Ruppert, Wand and Carroll (2003) and Proietti (2007)], and parametric methods based on the state space models or the Wiener–Kolmogorov signal extraction theory [Whittle (1983)]. An essential and up to date monograph on measuring trends and cycles in economics is Mills (2003). A problem that is common to all the methodologies is the reliability of the trend estimates at the end of the sample period.

The concern of the paper is real time estimation of the underlying trend in a time series by means of filters that arise from fitting a local polynomial of a given degree with a constant bandwidth. Real time estimation is of outmost importance in fields like economics and deals with the estimation of a signal at time $t$ using the observations available up to and including time $t$.

A well-known property of local polynomial estimators is the automatic adaptation at the boundaries. It essentially means that the bias and the variance near the boundary are of the same order of magnitude as in the interior. See, for instance, Hastie and Loader (1993), Fan and Gijbels (1996), Section 3.2.5, Simonoff (1996), Section 5.2.3 and the references therein.

It turns out, however, that for a local cubic fit, such as that arising from the well-known Henderson filter [Henderson (1916)], the variance inflation resulting from the one-sided real time direct filter is very high, and that the filter is strongly localized at the current observations, with a leverage that is close to unity.

The paper documents this basic feature and will be concerned, in particular, with the evaluation of alternative strategies aiming at the adaptation at the boundary of a given two-sided symmetric local polynomial filter. Our discussion will mostly refer to the Henderson filter. The latter has a long tradition for trend-cycle estimation in economic time series. The relevance of Henderson's contribution to modern local regression is stressed in the first chapter of Loader (1999). Henderson filters are still employed for trend estimation in the X-11 cascade filter, and as such are an integral part of the X-12-ARIMA procedure, the official seasonal adjustment procedure in the US, Canada, the UK and many other countries. See Dagum (1980), Findley et al. (1998) and Ladiray and Quenneville (2001) for more details.

The plan of the paper is the following. Section 2 motivates the applied problem of interest, presenting an example dealing with the assessment of recent business conditions in the US housing markets. After reviewing the constructive principles presiding the derivation of the two-sided symmetric local polynomial filters, Section 3 provides a thorough assessment of the properties of the asymmetric filters automatically adapted at the boundary, which result from fitting a local polynomial with a fixed bandwidth to the observations available at the current time. The direct asymmetric filters can be equivalently derived through the reproducing kernel Hilbert space method, as we prove based on the Hankel representation of a reproducing kernel in the context of weighted least squares estimation. The key result, as we stressed above, is that the real time filter behaves differently from the other automatically adapted asymmetric filters inside the boundary.

Section 4 evaluates an alternative class of fixed bandwidth asymmetric filters that result from minimizing the mean square revision error subject to polynomial reproduction constraints. This class generalizes the well-known Musgrave's asymmetric approximation of the Henderson filters [Musgrave (1964), see also Doherty (2001), Gray and Thomson (2002) and Quenneville, Ladiray and Lefrancois (2003)], which is implemented in the seasonal adjustment filter X-11, developed by the US Census Bureau [see Findley et al. (1998) and Ladiray and Quenneville (2001)]. The class of filters depends on the properties of the true underlying signal, namely, its level, slope, curvature and so forth, which can be estimated from the data.

In Section 5 we provide a few illustrations dealing with economic time series. They address the issue of approximating the Henderson filter in real time and show that the slope and curvature play a relevant role for the derivation of the optimal real time approximation. These features can be estimated from the available data. The two features are, on the contrary, neglected by Musgrave's asymmetric filters, which postulate that the true underlying signal is linear but only require that the approximate filter is capable of reproducing a zero degree polynomial. In Section 6 we draw our conclusions.

**2. A motivating example: assessing recent trends in the housing market.**
Figure 1 displays the monthly time series of *housing starts*, for the period

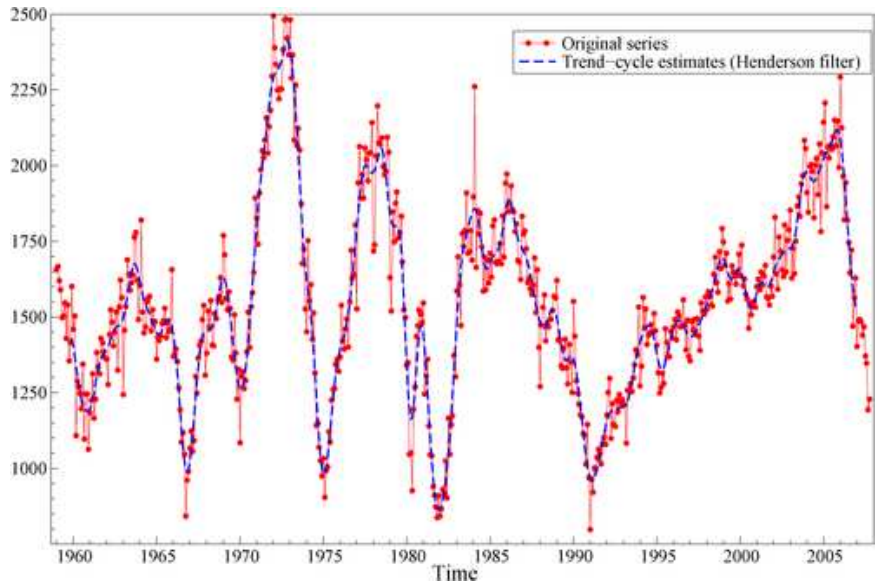

Fig. 1. *New Privately Owned Housing Units Started, US Source: Census Bureau. Original series and two-sided nonparametric trend estimates obtained by the Henderson filter.*

January 1959–October 2007. The series, published by the US Census Bureau, concerns the number of privately owned new housing units on which construction has been started over the reference period. See the US Census Bureau website at www.census.gov for further documentation.

Housing starts represent an important indicator of the state of the economy. In a recent paper Leamer (2007) argues that residential investment offers the best early warning sign of an oncoming recession. It is evident from Figure 1 that housing starts peaked at the end of 2005, and underwent thenceforth a very steep decline. The analyst is typically interested in the timely assessment of the turning point and of the most recent trends, in a noisy environment. The estimation of turning points typically requires a trend-cycle estimate, before the application of a dating algorithm, such as a Bry and Boschan (1971) routine, which is widely popular in economics. This operation is necessary in order to prevent high frequency fluctuations from interfering with the identification of turning points, producing many false candidates.

The dashed line overlaid to the plot is the nonparametric estimate of the trend-cycle component produced by local cubic regression using a particular kernel, the two-sided Henderson kernel, which is discussed in more detail and contextualized in the next section. The estimates are computed on 21 consecutive monthly observations: the observation at the time of interest and 10 observations on each side of it (the bandwidth has been selected by cross-validation). As pointed out in the introductory remarks, this is only one of the possible solutions to the signal extraction problem. Another possibility would be to estimate the component of interest by postulating a (semi)parametric model for it, for example, an integrated random walk, an ARIMA model or a smoothing spline. Be that as it may, the estimation at the end of the sample period is a delicate issue for any signal extraction method. Think, for instance, to wavelet multiresolution analysis: the traditional solution to the problem of handling boundary conditions, based on the circularity assumption [see Percival and Walden (2000), pages 197–199], is implausible here due to the nonstationary nature of the series.

Turning back our attention to our local polynomial approach, it is evident from the plot that the two-sided estimates of the signal are not available for the last 10 months, which are the most interesting from the point of view of the business cycle analyst and the policy maker. Actually, a direct solution is readily available: it arises from the automatic adaptation of a local cubic polynomial to the available observations at the end of the sample, using the same bandwidth (or a nearest neighbor bandwidth) and the same kernel. However, we shall argue in the paper that the corresponding estimates are inherently too volatile. This feature is visible from the plot of the real time estimates (i.e., the one-sided estimates using the current observation and 10 past observations) arising from the direct asymmetric adaptation, which are displayed in the left panel of Figure 2 (solid line) along with the original time series observations (dots) and the final two-sided Henderson filter estimates (dashed line). These estimates are very rough; they are close to the original observations and far away from the final two-sided estimates. As such, they potentially give rise to a large number of false candidate turning points. Moreover, the revision of the estimates as new information becomes available is substantial, as the comparison with the final Henderson estimates reveals.

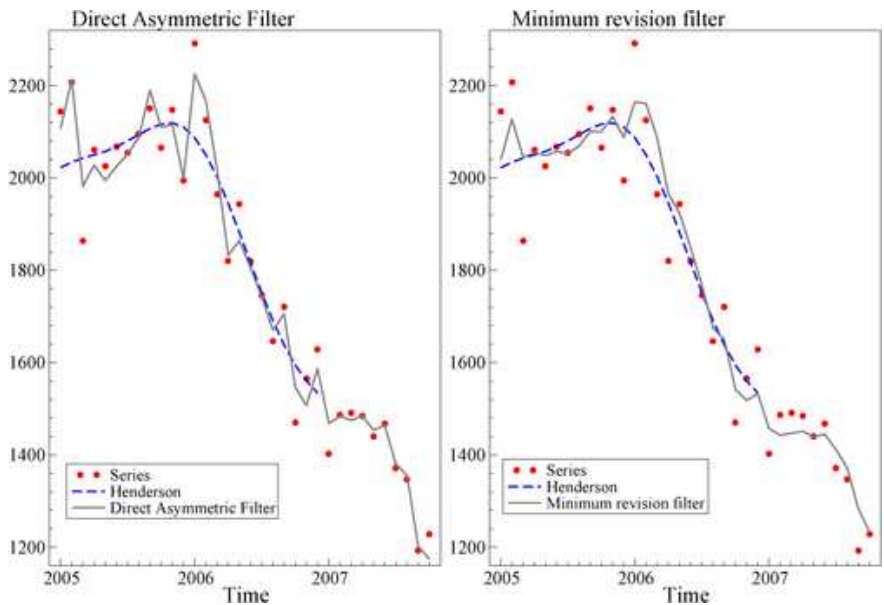

Fig. 2. *New Privately Owned Housing Units Started, US Source: Census Bureau. End of sample estimates of the trend-cycle component obtained by two asymmetric adaptations of the Henderson filter.*

The right panel presents the real time estimates produced by our proposed boundary filter. The plot reveals that they are more stable and more in line with the final ones. The proposed filter is derived according to the principle of minimizing the mean square revision error (the mean square deviation from the final Henderson estimates), subject to the condition that the asymmetric filter reproduces without distortion a linear trend and making the assumption that outside the sample period the underlying signal is a quadratic function of time (i.e., is a lower order polynomial outside the boundary of the sample space). Full details will be given in the sequel; it suffices to say at this point that we introduce bias, in order to reduce the variance of the estimates. This strategy proves effective for a large class of economic time series considered, as it will be illustrated further in Section 5.

We can draw here an analogy with the natural boundary conditions that are employed in cubic spline smoothing. Also in that framework, the nat-ural boundary conditions imply that the underlying signal behaves differently outside the boundary knots. In particular, the cubic spline is linear outside the boundary knots; see, for example, the discussion in Ruppert, Wand and Carroll (2003), page 72. In our case the assumption that the trend-cycle component has a lower order representation outside the sample period has similar statistical rationale, being designed to optimize the estimation bias-variance trade-off at the end of the sample space.

**3. Local polynomial filters and the Henderson filter.** Filters that arise from fitting a local polynomial have a well established tradition in time series analysis and signal extraction; see Anderson (1971), Chapter 3, Kendall (1973), Kendall, Stuart and Ord (1983), and the excellent historical review in Cleveland and Loader (1996). In this section we review the derivation of linear smoothers for trend extraction.

Let us assume that time is discrete and that the series can be decomposed as $y_t = \mu_t + \varepsilon_t$, where $\mu_t$ is the signal (trend) and $\varepsilon_t \sim \text{NID}(0, \sigma^2)$ is the noise. The signal is approximated locally by a polynomial of degree $d$, so that in the neighborhood of time $t$ we can write

$$y_{t+j} = m_{t+j} + \varepsilon_{t+j}, \qquad m_{t+j} = \beta_0 + \beta_1 j + \beta_2 j^2 + \cdots + \beta_d j^d,$$
$$j = 0, \pm 1, \ldots, \pm h.$$

In matrix notation, the local polynomial approximation can be written as follows:

$$\mathbf{y} = \mathbf{X}\boldsymbol{\beta} + \boldsymbol{\varepsilon}, \qquad \boldsymbol{\varepsilon} \sim \text{N}(\mathbf{0}, \sigma^2 \mathbf{I}), \tag{1}$$

where $\mathbf{y} = [y_{t-h}, \ldots, y_t, \ldots, y_{t+h}]'$, $\boldsymbol{\varepsilon} = [\varepsilon_{t-h}, \ldots, \varepsilon_t, \ldots, \varepsilon_{t+h}]'$,

$$\mathbf{X} = \begin{bmatrix} 1 & -h & h^2 & \vdots & (-h)^d \\ 1 & -(h-1) & (h-1)^2 & \vdots & [-(h-1)]^d \\ \vdots & \vdots & \cdots & \cdots & \vdots \\ 1 & 0 & 0 & \vdots & 0 \\ \vdots & \vdots & \cdots & \cdots & \vdots \\ 1 & h-1 & (h-1)^2 & \vdots & (h-1)^d \\ 1 & h & h^2 & \vdots & h^d \end{bmatrix}, \qquad \boldsymbol{\beta} = \begin{bmatrix} \beta_0 \\ \beta_1 \\ \vdots \\ \beta_d \end{bmatrix}.$$

Using this design, the value of the trend at time $t$ is simply given by the intercept, $m_t = \beta_0$. Provided that $2h \geq d$, the $d+1$ unknown coefficients $\beta_k$, $k = 0, \ldots, d$, can be estimated by the method of weighted least squares (WLS), which consists of minimizing with respect to the $\beta_k$'s the objective function:

$$S(\hat{\beta}_0, \ldots, \hat{\beta}_d) = \sum_{j=-h}^{h} \kappa_j (y_{t+j} - \hat{\beta}_0 - \hat{\beta}_1 j - \hat{\beta}_2 j^2 - \cdots - \hat{\beta}_d j^d)^2. \tag{2}$$

Here, $\kappa_j \geq 0$ is a set of weights that define, either explicitly or implicitly, a kernel function.

Reverting to the matrix notation, setting $\mathbf{K} = \mathrm{diag}(\kappa_{-h}, \ldots, \kappa_{-1}, \kappa_0, \kappa_1, \ldots, \kappa_h)$, the WLS estimate of the coefficients is $\hat{\boldsymbol{\beta}} = (\mathbf{X}'\mathbf{K}\mathbf{X})^{-1}\mathbf{X}'\mathbf{K}\mathbf{y}$. In order to obtain $\hat{m}_t = \hat{\beta}_0$, we need to select the first element of the vector $\hat{\boldsymbol{\beta}}$. Hence, denoting by $\mathbf{e}_1$ the $d+1$ vector $\mathbf{e}_1' = [1, 0, \ldots, 0]$,

$$\hat{m}_t = \mathbf{e}_1'\hat{\boldsymbol{\beta}} = \mathbf{e}_1'(\mathbf{X}'\mathbf{K}\mathbf{X})^{-1}\mathbf{X}'\mathbf{K}\mathbf{y} = \mathbf{w}'\mathbf{y} = \sum_{j=-h}^{h} \mathrm{w}_j y_{t-j},$$

which expresses the estimate of the trend as a linear combination of the observations with coefficients

(3) $$\mathbf{w} = \mathbf{K}\mathbf{X}(\mathbf{X}'\mathbf{K}\mathbf{X})^{-1}\mathbf{e}_1.$$

The linear combination yielding the trend estimate is the local polynomial two-sided *filter*. It satisfies $\mathbf{X}'\mathbf{w} = \mathbf{e}_1$, or equivalently,

$$\sum_{j=-h}^{h} \mathrm{w}_j = 1, \qquad \sum_{j=-h}^{h} j^r \mathrm{w}_j = 0, \qquad r = 1, 2, \ldots, d.$$

As a consequence, the filter $\mathbf{w}$ is said to preserve a deterministic polynomial of order $d$. Moreover, the filter weights are symmetric ($\mathrm{w}_j = \mathrm{w}_{-j}$), which follows from the symmetry of the kernel weights $\kappa_j$, and the assumption that the available observations are equally spaced.

The Henderson filter [see Henderson (1916), Kenny and Durbin (1982), Loader (1999), Ladiray and Quenneville (2001)] arises as the weighted least squares estimator of a local cubic trend at time $t$ using $2h+1$ consecutive observations. When $d = 3$, the weights in (3) take the form

$$\mathrm{w}_j = \kappa_j \frac{(S_4 - S_2 j^2)}{S_0 S_4 - S_2^2}, \qquad j = 0, \pm 1, \ldots, \pm h,$$

where $S_r = \sum_{j=-h}^{h} \kappa_j j^r$. This expression makes the dependence on the kernel weights, $\kappa_j$, explicit. Henderson (1916) addressed the problem of defining a set of kernel weights that maximize the smoothness of the estimated trend, in the sense that the variance of its third differences is as small as possible. He showed that up to a factor of proportionality we must have $\kappa_j = [(h+1)^2 - j^2][(h+2)^2 - j^2][(h+3)^2 - j^2]$. Hence, the coefficients $\kappa_j$ given above define the (unnormalized) Henderson kernel. It can be shown that $\kappa_j$ minimize the sum of squared third order differences of the weights sequence, $\mathrm{w}_j$.

3.1. *Asymmetric filters and their automatic adaptation at boundary points.* The derivation of the two-sided symmetric filter has assumed the availability of $2h+1$ observations centered at $t$. Obviously, for a given finite sequence $y_t, t = 1, \ldots, n$, it is not possible to obtain the estimates of the signal for the (first and) last $h$ time points, which is inconvenient, since we are typically most interested at the most recent estimates.

We can envisage three fundamental approaches to the estimation of the signal at the extremes of the sample period:

1. The construction of asymmetric filters that result from fitting a local polynomial to the available observations $y_t, t = n-h+1, n-h+2, \ldots, n$.
2. Apply the symmetric two-sided filter **w** to the series extended by $h$ forecasts $\hat{y}_{n+l|n}, l = 1, \ldots, h$ (and backcasts $\hat{y}_{1-l|n}$).
3. Derive the best approximating filter which minimizes the revision mean square error subject to polynomial reproducing constraints.

The second strategy (using forecast extensions) is safer, provided that we are capable of producing optimal forecasts according to some parametric or nonparametric device, for example, by fitting a time series model of the ARIMA class. This idea is embodied in the X-11-ARIMA seasonal adjustment procedure [Dagum (1982)]. An intuitive and easily established fact is that if the forecasts $\hat{y}_{n+l|n}$ are optimal in the mean square error sense, then the variance of the revision is a minimum; see Wallis (1983). In applied economic time series analysis most often extrapolations have a local linear nature, such as those obtained from ARIMA models with integration order equal to 1 or 2 (provided there is no constant term in the latter case). Recently, Dagum and Luati (2009) derived linear asymmetric filters based on data independent extrapolations from fixed ARIMA models and parameter values. When the forecast extensions are exogenous, the filter weights are adapted to the property of the series, so that the weights $w_j$ are not fixed, but depend also on the ARIMA model for $y_t$.

The trend estimates for the last $h$ data points, $\hat{m}_{n-h+1|n}, \ldots, \hat{m}_{n|n}$, use respectively $2h, 2h-1, \ldots, h+1$ observations. It is thus inevitable that the last $h$ estimates of the trend will be subject to revision as new observations become available. In the sequel we shall denote by $q$ the number of future observations available at time $t$ (the period which our estimate is referred to), $q = 0, \ldots, h$, and by $\hat{m}_{t|t+q}$ the estimate of the signal at time $t$ using the information available up to time $t+q$, with $0 \leq q \leq h$; $\hat{m}_{t|t}$ is usually known as the real time estimate since it uses only the past and current information.

We now deal with the first strategy, which results from the automatic adaptation of the local polynomial filter to the available sample; we then interpret the results in terms of the other two strategies. The approximate model $y_{t+j} = m_{t+j} + \varepsilon_{t+j}$ is assumed to hold for $j = -h, -h+1, \ldots, q$, and the estimators of the coefficients $\hat{\beta}_k$, $k=0,\ldots,d$, minimize

$$S(\hat{\beta}_0,\ldots,\hat{\beta}_d) = \sum_{j=-h}^{q} \kappa_j(y_{t+j} - \hat{\beta}_0 - \hat{\beta}_1 j - \hat{\beta}_2 j^2 - \cdots - \hat{\beta}_d j^d)^2.$$

Let us partition the matrices $\mathbf{X}$, $\mathbf{K}$ and the vector $\mathbf{y}$ as follows:

$$\mathbf{X} = \begin{bmatrix} \mathbf{X}_p \\ \mathbf{X}_f \end{bmatrix}, \qquad \mathbf{y} = \begin{bmatrix} \mathbf{y}_p \\ \mathbf{y}_f \end{bmatrix}, \qquad \mathbf{K} = \begin{bmatrix} \mathbf{K}_p & \mathbf{0} \\ \mathbf{0} & \mathbf{K}_f \end{bmatrix},$$

where $\mathbf{y}_p$ denotes the set of available observations, whereas $\mathbf{y}_f$ is missing and $\mathbf{X}$ and $\mathbf{K}$ are partitioned accordingly. The direct asymmetric filter (DAF) arising as the solution to the above weighted least squares problem is written in matrix notation as

(4) $$\mathbf{w}_a = \mathbf{K}_p \mathbf{X}_p (\mathbf{X}_p' \mathbf{K}_p \mathbf{X}_p)^{-1} \mathbf{e}_1.$$

The filter resulting from the automatic adaptation of the local polynomial fit can be equivalently derived using the second strategy, assuming that the future observations are generated according to a polynomial function of time of degree $d$, so that the optimal forecasts are generated by the same polynomial model. Under the local polynomial model the forecasted values of $\mathbf{y}_f$ are

$$\hat{\mathbf{y}}_f = \mathbf{X}_f (\mathbf{X}_p' \mathbf{K}_p \mathbf{X}_p)^{-1} \mathbf{X}_p' \mathbf{K}_p \mathbf{y}_p.$$

Applying the two-sided filter $\mathbf{w}$ to the observations extended by the forecasts yields

$$\hat{m}_{t|t+q} = \mathbf{w}' \begin{bmatrix} \mathbf{y}_p \\ \hat{\mathbf{y}}_f \end{bmatrix} = \mathbf{e}_1' (\mathbf{X}'\mathbf{K}\mathbf{X})^{-1} \mathbf{X}'\mathbf{K} \begin{bmatrix} \mathbf{y}_p \\ \hat{\mathbf{y}}_f \end{bmatrix};$$

using $\mathbf{X}'\mathbf{K} = [\mathbf{X}_p' \mathbf{K}_p, \mathbf{X}_f' \mathbf{K}_f]$,

$$\begin{aligned}(\mathbf{X}'\mathbf{K}\mathbf{X})^{-1} &= (\mathbf{X}_p' \mathbf{K}_p \mathbf{X}_p + \mathbf{X}_f' \mathbf{K}_f \mathbf{X}_f)^{-1} \\ &= (\mathbf{X}_p' \mathbf{K}_p \mathbf{X}_p)^{-1}[\mathbf{I} + \mathbf{X}_f' \mathbf{K}_f \mathbf{X}_f (\mathbf{X}_p' \mathbf{K}_p \mathbf{X}_p)^{-1}]^{-1}\end{aligned}$$

and replacing $\hat{\mathbf{y}}_f$ gives

$$\hat{m}_{t|t+q} = \mathbf{e}_1' (\mathbf{X}_p' \mathbf{K}_p \mathbf{X}_p)^{-1} \mathbf{X}_p' \mathbf{K}_p \mathbf{y}_p,$$

which is also the estimate of the intercept of the polynomial that uses only the available information. Hence, the asymmetric filter weights that are automatically adapted at the boundaries are given by (4).

The explicit expressions for the weights of the DAF, $\mathbf{w}_a$, are derived below for $d \leq 3$, based on (4) and on matrix inversion formulae. Setting $S_{qr} =$

$\sum_{j=-h}^{q} j^r \kappa_j$ for $q = 0, \ldots, h$, the solutions for $d = 0, 1, 2$ are, respectively,

$$w_{a,j} = \frac{\kappa_j}{S_{q0}}, \qquad w_{a,j} = \kappa_j \frac{S_{q2} - jS_{q1}}{S_{q0}S_{q2} - S_{q1}^2},$$

$$w_{a,j} = \kappa_j \frac{\zeta_{4,2} - j\zeta_{4,1} + j^2\zeta_{3,1}}{S_{q0}\zeta_{4,2} - S_{q1}\zeta_{4,1} + S_{q2}\zeta_{3,1}},$$

$j = -q, \ldots, h$, where $\zeta_{m,n} = S_{qm}S_{qn} - S_{q,m-1}S_{q,n-1}$. For $d = 3$,

(5) $\qquad w_{a,j} = \kappa_j \frac{Z_0 - Z_1 j + Z_2 j^2 - Z_3 j^3}{S_{q0}Z_0 - S_{q1}Z_1 + S_{q2}Z_2 - S_{q3}Z_3}, \qquad j = -q, \ldots, h,$

where $Z_0 = S_{q2}\zeta_{6,4} - S_{q3}\zeta_{6,3} + S_{q4}\zeta_{5,3}$, $Z_1 = -(S_{q1}\zeta_{6,4} - S_{q2}\zeta_{6,3} + S_{q3}\zeta_{5,3})$, $Z_2 = S_{q1}\zeta_{6,3} - S_{q2}\zeta_{6,2} + S_{q4}\zeta_{4,2}$, $Z_3 = -(S_{q1}\zeta_{5,3} - S_{q2}\zeta_{5,2} + S_{q3}\zeta_{4,2})$.

The real time filters arise when $q = 0$ in the above expressions. The symmetric weights of the smoothing filter, $\mathbf{w}$, arise instead when $q = h$ in the above expressions. Replacing $S_{hr} = 0$ for $r$ odd, we find

$$w_j = \frac{\kappa_j}{S_{h0}}, \qquad w_j = \kappa_j \frac{S_{h4} - j^2 S_{h2}}{S_{h0}S_{h4} - S_{h2}^2}$$

for $d = 0, 1$, and $d = 2, 3$, respectively.

The direct asymmetric filters can be alternatively derived with the reproducing kernel Hilbert space (RKHS) approach [see Berlinet and Thomas-Agnan (2004)]. In that context, the equivalent kernel of a linear estimator of order $d$ can be obtained as $K_d(t) = R_d(t, 0)f_0(t)$, where $R_d(t, 0)$ is the reproducing kernel of a Hilbert space of polynomials up to degree $d \geq 1$ with inner product defined with respect to a density function $f_0(t)$. The reproducing kernel is so-called because it reproduces any function in the Hilbert space in the sense that $\langle g, R_d(t, \cdot) \rangle_{\mathcal{H}} = g(t)$, $\forall t \in T$, $g \in \mathcal{H}$, from which many inferential properties can be derived. Once $f_0(t)$ is chosen with finite moments $\nu_0, \nu_1, \ldots, \nu_{2d}$, one way to obtain the associated reproducing kernel is by means of Hankel determinants [Berlinet and Thomas-Agnan (2004), Theorem 80], in that

$$K_d(t) = \frac{\det(\mathbf{H}_d^0[1, \mathbf{x}_t])}{\det(\mathbf{H}_d^0)} f_0(t),$$

where $\mathbf{H}_d^0$ is the Hankel matrix whose elements are the moments of $f_0(t)$, from $\nu_0$ to $\nu_d$ in the first row and from $\nu_d$ to $\nu_{2d}$ in the last column, and $\mathbf{H}_d^0[1, \mathbf{x}_t]$ is the matrix obtained replacing the first column of $\mathbf{H}_d^0$ by the vector $\mathbf{x}_t = [1, t, t^2, \ldots, t^d]'$. In our discrete setting, choosing the (normalized) Henderson kernel $\kappa_j$ in place of the density $f_0(t)$, then $\nu_r = S_{qr}$ for $r =$ $0, \ldots, 2d$ and the matrix $\mathbf{H}_d^0$ becomes $\mathbf{X}_p' \mathbf{K}_p \mathbf{X}_p$, so that the elements of the filter $\mathbf{w}_a$ are given by

$$\text{(6)} \qquad \mathrm{w}_{a,j} = \frac{\det(\mathbf{X}_p' \mathbf{K}_p \mathbf{X}_p[1, \mathbf{x}_j])}{\det(\mathbf{X}_p' \mathbf{K}_p \mathbf{X}_p)} \kappa_j,$$

where $\mathbf{x}_j = [1, j, j^2, \ldots, j^d]'$. The expression (6) is exactly the same that we would obtain by solving for $\hat{\beta}_0 = \hat{m}_t$ the least squares equation

$$(\mathbf{X}_p' \mathbf{K}_p \mathbf{X}_p) \hat{\boldsymbol{\beta}} = \mathbf{X}_p' \mathbf{K}_p \mathbf{y}_p,$$

using the Cramer rule for the explicit solution of a linear system. In fact, setting $\mathbf{b} = \mathbf{X}_p' \mathbf{K}_p \mathbf{y}_p$, the first coordinate of the solution vector is

$$\hat{\beta}_0 = \frac{\det(\mathbf{X}_p' \mathbf{K}_p \mathbf{X}_p[1, \mathbf{b}])}{\det(\mathbf{X}_p' \mathbf{K}_p \mathbf{X}_p)}.$$

Given that $\mathbf{b} = \sum_{j=-h}^{q} \mathbf{x}_j \kappa_j y_{t+j}$, then

$$\det(\mathbf{X}_p' \mathbf{K}_p \mathbf{X}_p[1, \mathbf{b}]) = \sum_{j=-h}^{q} \det(\mathbf{X}_p' \mathbf{K}_p \mathbf{X}_p[1, \mathbf{x}_j]) \kappa_j y_{t+j}$$

and, therefore,

$$\hat{m}_t = \sum_{j=-h}^{q} \frac{\det(\mathbf{X}_p' \mathbf{K}_p \mathbf{X}_p[1, \mathbf{x}_j])}{\det(\mathbf{X}_p' \mathbf{K}_p \mathbf{X}_p)} \kappa_j y_{t+j},$$

from which (6) follows.

The above expression also holds for symmetric filters, arising when $q = h$, and for any choice of the kernel $\kappa_j$, providing an alternative way to express both the trend estimate and the equivalent kernel of the linear filter resulting by weighted linear regression.

3.2. *Properties of the direct asymmetric filters.* Let us partition the weights of the two-sided symmetric filter in two groups, $\mathbf{w} = [\mathbf{w}_p', \mathbf{w}_f']'$, where $\mathbf{w}_p$ contains the weights attributed to the past and current observations and $\mathbf{w}_f$ those attached to the future unavailable observations. Then,

$$\begin{aligned}
\mathbf{w}_p &= \mathbf{K}_p \mathbf{X}_p (\mathbf{X}' \mathbf{K} \mathbf{X})^{-1} \mathbf{e}_1 \\
&= \mathbf{K}_p \mathbf{X}_p (\mathbf{X}_p' \mathbf{K}_p \mathbf{X}_p + \mathbf{X}_f' \mathbf{K}_f \mathbf{X}_f)^{-1} \mathbf{e}_1 \\
&= [\mathbf{K}_p \mathbf{X}_p (\mathbf{X}_p' \mathbf{K}_p \mathbf{X}_p)^{-1} \\
&\quad - \mathbf{K}_p \mathbf{X}_p (\mathbf{X}_p' \mathbf{K}_p \mathbf{X}_p)^{-1} \mathbf{X}_f' \mathbf{K}_f \mathbf{X}_f (\mathbf{X}_p' \mathbf{K}_p \mathbf{X}_p + \mathbf{X}_f' \mathbf{K}_f \mathbf{X}_f)^{-1}] \mathbf{e}_1 \\
&= \mathbf{w}_a - \mathbf{K}_p \mathbf{X}_p (\mathbf{X}_p' \mathbf{K}_p \mathbf{X}_p)^{-1} \mathbf{X}_f' \mathbf{w}_f,
\end{aligned}$$

as $\mathbf{w}_f = \mathbf{K}_f \mathbf{X}_f (\mathbf{X}'_p \mathbf{K}_p \mathbf{X}_p + \mathbf{X}'_f \mathbf{K}_f \mathbf{X}_f)^{-1} \mathbf{e}_1$.

Thus, we have the fundamental relationship which states how the asymmetric filter weights are obtained from the symmetric ones:

$$\mathbf{w}_a = \mathbf{w}_p + \mathbf{K}_p \mathbf{X}_p (\mathbf{X}'_p \mathbf{K}_p \mathbf{X}_p)^{-1} \mathbf{X}'_f \mathbf{w}_f. \tag{7}$$

Premultiplying both sides by $\mathbf{X}'_p$, we can see that the asymmetric filter weights satisfy the following polynomial reproduction constraints:

$$\mathbf{X}'_p \mathbf{w}_a = \mathbf{X}'_p \mathbf{w}_p + \mathbf{X}'_f \mathbf{w}_f = \mathbf{X}' \mathbf{w}.$$

If the design of the time points is centered around the current time, then $\mathbf{X}'\mathbf{w} = \mathbf{e}_1$. Thus, the bias in estimating an unknown function of time has the same order of magnitude as in the interior of time support.

We now show that the weights $\mathbf{w}_a$ are the unique minimizers with respect to $\mathbf{v}$ of the following constrained problem:

$$\min_{\mathbf{v}} (\mathbf{v} - \mathbf{w}_p)' \mathbf{K}_p^{-1} (\mathbf{v} - \mathbf{w}_p) \qquad \text{s.t. } \mathbf{X}'_p \mathbf{v} = \mathbf{X}' \mathbf{w},$$

where $\mathbf{w} = [\mathbf{w}'_p, \mathbf{w}'_f]'$. The first order conditions give $\mathbf{v} = \mathbf{w}_p + \mathbf{K}_p \mathbf{X}_p \boldsymbol{l}$, where $\boldsymbol{l}$ is a vector of Lagrange multipliers. Premultiplying both sides by $\mathbf{X}'_p$ and replacing $\mathbf{X}'_p \mathbf{v} = \mathbf{X}' \mathbf{w}$ gives $\boldsymbol{l} = (\mathbf{X}'_p \mathbf{K}_p \mathbf{X}_p)^{-1} \mathbf{X}'_f \mathbf{w}_f$, and replacing into the expression for $\mathbf{v}$ gives $\mathbf{v} = \mathbf{w}_a$, as given by (7).

Hence, the asymmetric weights $\mathbf{w}_a$ minimize the weighted distance between the asymmetric filter coefficients and the symmetric ones, where the weights are provided by the reciprocal of the kernel weights. This result is useful in order to compare the asymmetric direct filter with the class of asymmetric filters derived in Section 4.

Figure 3 plots the weights of the direct asymmetric Henderson filter for $q$ ranging from 0 (real time filter) to $h$ (symmetric Henderson filter), along with their gain when the bandwidth takes the value $h = 6$, producing the Henderson 13 terms moving average when all the necessary observations are available. The real time filter uses 7 consecutive observations and it is very much concentrated on the current observation. As a consequence, the gain behaves rather poorly, being close to one also at the high frequencies.

Hence, our analysis reveals a sort of discontinuity in the behavior of the filter, when we move from $q = 0$ (real time filter) to $q = 1$ (one future observation is available). The real time filter is unbiased if the series is generated by a cubic polynomial; however, the preservation of the bias properties is done at the expenses of the variance, which is very high, since most of the contribution to the trend estimate comes from the current observation. This can be explained by means of the following relation, that gives the leverage of the filter, that is, the weight attached to the observation taken at the same time we are interested in the trend estimate,

$$\mathrm{w}_{a,0} = \kappa_0 \mathbf{e}'_1 (\mathbf{X}'_p \mathbf{K}_p \mathbf{X}_p)^{-1} \mathbf{e}_1 = \kappa_0 \frac{\det(\mathbf{M}_{1,1})}{\det(\mathbf{X}'_p \mathbf{K}_p \mathbf{X}_p)},$$

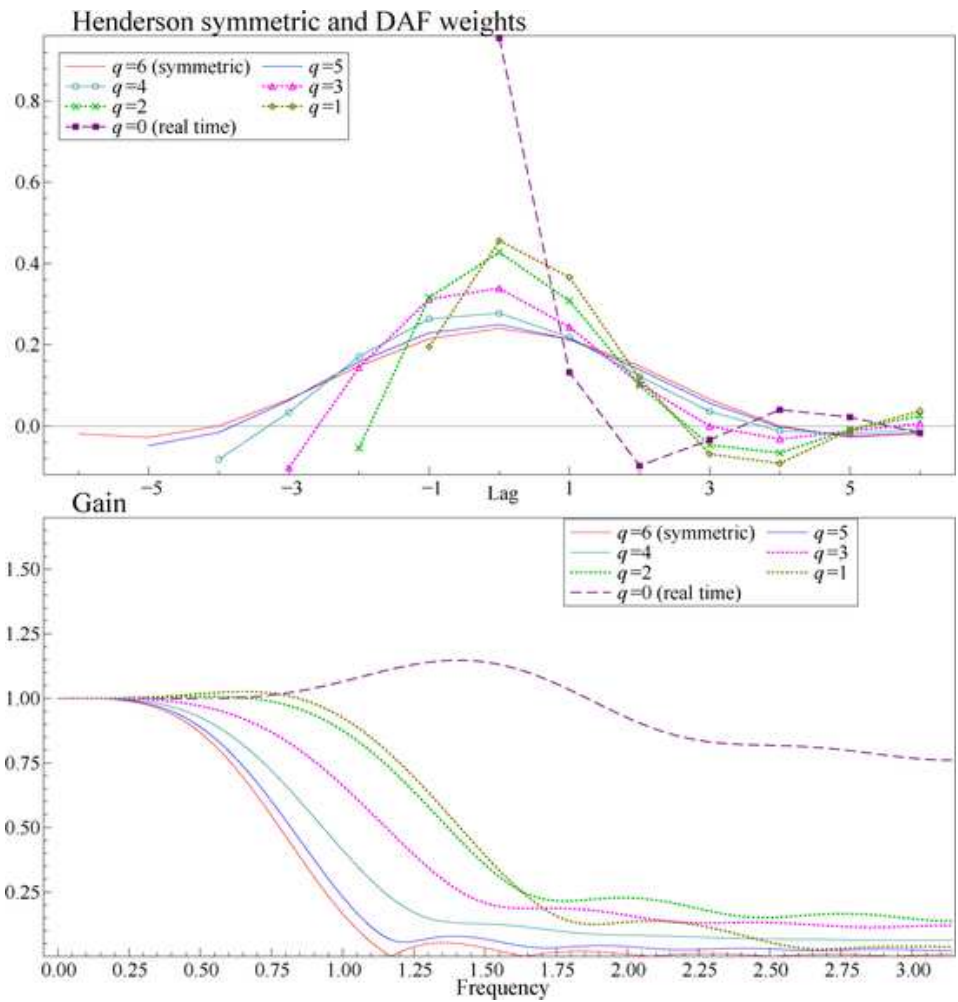

FIG. 3. *Gain, phase and weights for the symmetric and asymmetric Henderson filters* $\mathbf{w}_a$; $q$ *is the number of future observations available for estimating the signal.*

where $\mathbf{M}_{1,1}$ is the submatrix obtained by deleting the first row and column of $\mathbf{X}'_p \mathbf{K}_p \mathbf{X}_p$.

(i) For fixed values of $d$, the leverage decreases as long as the span of the filter increases. It is maximum for the real time filter ($q=0$) and minimum for the symmetric filter ($q=h$).

(ii) On the other hand, for fixed values of $h$ or $q$ the leverage exponentially increases if the degree of the fitting polynomial increases. It is minimum for $d=0$ and maximum for $d=h$.

In particular, $w_{a,0} = 1$ for $d = h$. The latter equality can be proved by noticing the general fact that the $h+1$th row of $\mathbf{X}_p$ (last row when real time filters are considered), whose elements correspond to $j^r$, $r = 0, \ldots, d$, is the vector $\mathbf{e}'_1 = [1, 0, 0, \ldots, 0]$. Given that $\mathbf{K}_p$ is diagonal, it follows from the row-column matrix product that

$$\mathbf{M}_{1,1} = \mathbf{X}'_{h+1,1} \mathbf{K}_{h+1,h+1} \mathbf{X}_{h+1,1},$$

where $\mathbf{X}_{i,j}$ and $\mathbf{K}_{i,j}$ are submatrices obtained by deleting the $i$th row and $j$th column of $\mathbf{X}_p$ and $\mathbf{K}_p$. If $d = h$, then $\mathbf{X}_p$ and $\mathbf{X}_{h+1,1}$ are square matrices that have different dimensions but the same determinant, as it is immediate to see by calculating $\det(\mathbf{X}_p)$ from the last row of $\mathbf{X}_p$ with the Laplace formula. Hence, it follows from the Binet–Cauchy theorem that

$$\frac{\det(\mathbf{M}_{1,1})}{\det(\mathbf{X}'_p \mathbf{K}_p \mathbf{X}_p)} = \frac{\det(\mathbf{X}'_{h+1,1})\det(\mathbf{K}_{h+1,h+1})\det(\mathbf{X}_{h+1,1})}{\det(\mathbf{X}'_p)\det(\mathbf{K}_p)\det(\mathbf{X}_p)} = \frac{1}{\kappa_0}$$

and, therefore, $w_{a,0} = 1$. Since the filter reproduces polynomials up to the order $d$, $w_{a,j} = 0$ for $j = -h, \ldots, -1$. This result holds also for symmetric and nearest neighbor filters, where the maximum value $d$ can assume is $2h$. Proof of (i) and (ii) is given in the Appendix based on a generalized version of the Binet–Cauchy theorem. Note that these relations can be verified by (5), up to $d = 3$, and by (6).

Table 1 illustrates how $w_{a,0}$ varies with the length of the asymmetric filters and the degree of the fitting polynomial. The values are calculated for $h = 6$ and $d$ ranging from $d = 0$ (constant trend) to $d = 6$ (six degree polynomial) and $q$ ranging from $q = 0$ (real time filter) to $q = 6$, which gives the symmetric 13 term Henderson filter. It is evident, and not surprising, that the impact, on the leverage, of the degree of the fitting polynomial is much greater than that of the span of the filter. Even for small values of the order of the approximating polynomial, the leverage of the real time filter results in greater than 0.5.

We have alternatively evaluated the values of Table 1 using different kernels, such as the Uniform and the Epanechnikov [Epanechnikov (1969)], but the resulting real time filters are almost equivalent to those calculated with the Henderson kernel, and therefore not illustrated here.

**4. On a general class of asymmetric filters.** We now consider a class of asymmetric filters approximating a given symmetric two-sided smoothing filter. The class depends on unknown features of the series, such as slope and curvature, which can be estimated from the data, and encompasses the so-called Musgrave's surrogate filters (1964). The latter, which will be discussed in Section 4.1, approximate the two-sided Henderson filter at the end of the sample and are a component of the well-known X-11 cascade seasonal adjustment filter.

Table 1
*Values of* $w_{a,0}$ *for* $h = 6$, *different orders* $d$ *of the local polynomial and different values of* $q$ *ranging from* $q = 0$ *(real time) to* $q = 6$ *(symmetric filter)*

|  | $d = 0$ | $d = 1$ | $d = 2$ | $d = 3$ | $d = 4$ | $d = 5$ | $d = 6$ |
|---|---|---|---|---|---|---|---|
| $q = 0$ | 0.2457 | 0.5856 | 0.8356 | 0.9552 | 0.9925 | 0.9994 | 1.0000 |
| $q = 1$ | 0.1991 | 0.3038 | 0.3060 | 0.4560 | 0.7285 | 0.9238 | 0.9908 |
| $q = 2$ | 0.1712 | 0.2008 | 0.2653 | 0.4275 | 0.4493 | 0.5189 | 0.7662 |
| $q = 3$ | 0.1547 | 0.1615 | 0.2652 | 0.3385 | 0.3603 | 0.5144 | 0.5397 |
| $q = 4$ | 0.1456 | 0.1466 | 0.2578 | 0.2776 | 0.3577 | 0.4309 | 0.4594 |
| $q = 5$ | 0.1413 | 0.1414 | 0.2472 | 0.2495 | 0.3516 | 0.3644 | 0.4593 |
| $q = 6$ | 0.1400 | 0.1400 | 0.2400 | 0.2400 | 0.3379 | 0.3379 | 0.4418 |

The minimum mean square revision error strategy which is at the basis of the criterion (8) was originally proposed by Musgrave (1964). Gray and Thomson (2002) generalized this idea to the case of a series generated by a local dynamic model. In this section we propose a different derivation of Gray and Thomson's result that is more general and clarifies some issues of the design of asymmetric filters, among which the connections with the DAF. We also provide an alternative expression for the asymmetric weights that is directly connected to Musgrave's result.

Assume that the observations are generated as $\mathbf{y} = \mathbf{U}\boldsymbol{\gamma} + \mathbf{Z}\boldsymbol{\delta} + \boldsymbol{\varepsilon}$, $\boldsymbol{\varepsilon} \sim N(\mathbf{0}, \mathbf{D})$, where $\mathbf{U}$, $\mathbf{Z}$ are a suitable design matrix. We aim at determining the asymmetric filter $\mathbf{v}$ minimizing the mean square revision error subject to constraints. The constraints are specified as follows: $\mathbf{U}'_p \mathbf{v} = \mathbf{U}'\mathbf{w}$, where $\mathbf{U} = [\mathbf{U}'_p, \mathbf{U}'_f]'$. Assuming that $[\mathbf{U}, \mathbf{Z}]$ is full column rank (usually, as it will be illustrated later, $[\mathbf{U}, \mathbf{Z}]$ is a selection of the columns of $\mathbf{X}$ or it is coincident with $\mathbf{X}$), and partitioning $\mathbf{D} = \mathrm{diag}(\mathbf{D}_p, \mathbf{D}_f)$, the set of asymmetric weights minimizes with respect to $\mathbf{v}$ the following objective function:

$$
\varphi(\mathbf{v}) = (\mathbf{v} - \mathbf{w}_p)' \mathbf{D}_p (\mathbf{v} - \mathbf{w}_p) + \mathbf{w}'_f \mathbf{D}_f \mathbf{w}_f \\
+ [\boldsymbol{\delta}'(\mathbf{Z}'_p \mathbf{v} - \mathbf{Z}'\mathbf{w})]^2 + 2\boldsymbol{l}'(\mathbf{U}'_p \mathbf{v} - \mathbf{U}'\mathbf{w}). \tag{8}
$$

The revision error arising in estimating the signal $m_t$ is $\hat{m}_{t|t} - \hat{m}_t = \mathbf{v}'\mathbf{y}_p - \mathbf{w}'\mathbf{y}$. Replacing $\mathbf{y}_p = \mathbf{U}_p\boldsymbol{\gamma} + \mathbf{Z}_p\boldsymbol{\delta} + \boldsymbol{\varepsilon}_p$, and $\mathbf{y} = \mathbf{U}\boldsymbol{\gamma} + \mathbf{Z}\boldsymbol{\delta} + \boldsymbol{\varepsilon}$, and using $\mathbf{U}'_p \mathbf{v} - \mathbf{U}'\mathbf{w} = \mathbf{0}$, we obtain $\hat{m}_{t|t} - \hat{m}_t = (\mathbf{v}'\mathbf{Z}_p - \mathbf{w}'\mathbf{Z})\boldsymbol{\delta} + \mathbf{v}'\boldsymbol{\varepsilon}_p - \mathbf{w}'\boldsymbol{\varepsilon}$, where $\boldsymbol{\varepsilon} = [\boldsymbol{\varepsilon}'_p, \boldsymbol{\varepsilon}'_f]'$. Hence, the first three summands of (8) represent the mean square revision error, which is broken down into the revision error variance (the first two terms) and the squared bias term $[\boldsymbol{\delta}'(\mathbf{Z}'_p \mathbf{v} - \mathbf{Z}'\mathbf{w})]^2$. The vector $\boldsymbol{l}$ is a vector of Lagrange multipliers.

Setting

$$\mathbf{Q} = \mathbf{D}_p + \mathbf{Z}_p \boldsymbol{\delta}\boldsymbol{\delta}' \mathbf{Z}'_p,$$

the first order conditions for the minimization problem can be written as follows:

$$\mathbf{v} = \mathbf{w}_p + \mathbf{Q}^{-1}\mathbf{Z}_p\boldsymbol{\delta}\boldsymbol{\delta}'\mathbf{Z}_f'\mathbf{w}_f - \mathbf{Q}^{-1}\mathbf{U}_p\boldsymbol{l}.$$

Premultiplying both sides for $\mathbf{U}_p'$ and recalling $\mathbf{U}_p'(\mathbf{v} - \mathbf{w}_p) = \mathbf{U}_f'\mathbf{w}_f$,

$$\mathbf{U}_f'\mathbf{w}_f = \mathbf{U}_p'\mathbf{Q}^{-1}\mathbf{Z}_p\boldsymbol{\delta}\boldsymbol{\delta}'\mathbf{Z}_f'\mathbf{w}_f - \mathbf{U}_p'\mathbf{Q}^{-1}\mathbf{U}_p\boldsymbol{l},$$

we can express the Lagrange multipliers as a linear combination of the weights $\mathbf{w}_f$:

$$\boldsymbol{l} = -[\mathbf{U}_p'\mathbf{Q}^{-1}\mathbf{U}_p]^{-1}[\mathbf{U}_f' - \mathbf{U}_p'\mathbf{Q}^{-1}\mathbf{Z}_p\boldsymbol{\delta}\boldsymbol{\delta}'\mathbf{Z}_f']\mathbf{w}_f.$$

Replacing into the expression for $\mathbf{v}$ yields

$$\mathbf{v} = \mathbf{w}_p + \mathbf{Q}^{-1}\mathbf{Z}_p\boldsymbol{\delta}\boldsymbol{\delta}'\mathbf{Z}_f\mathbf{w}_f + \mathbf{Q}^{-1}[\mathbf{U}_p'\mathbf{Q}^{-1}\mathbf{U}_p]^{-1}[\mathbf{U}_f' - \mathbf{U}_p'\mathbf{Q}^{-1}\mathbf{Z}_p\boldsymbol{\delta}\boldsymbol{\delta}'\mathbf{Z}_f']\mathbf{w}_f,$$

and rearranging,

$$(9) \qquad \mathbf{v} = \mathbf{w}_p + \mathbf{L}\mathbf{U}_f'\mathbf{w}_f + \mathbf{M}\mathbf{Z}_p\boldsymbol{\delta}\boldsymbol{\delta}'\mathbf{Z}_f'\mathbf{w}_f,$$

with

$$\mathbf{M} = \mathbf{Q}^{-1} - \mathbf{Q}^{-1}\mathbf{U}_p[\mathbf{U}_p'\mathbf{Q}^{-1}\mathbf{U}_p]^{-1}\mathbf{U}_p'\mathbf{Q}^{-1}, \qquad \mathbf{L} = \mathbf{Q}^{-1}\mathbf{U}_p[\mathbf{U}_p'\mathbf{Q}^{-1}\mathbf{U}_p]^{-1}.$$

The matrices $\mathbf{M}$ and $\mathbf{L}$ have the following properties: $\mathbf{U}_p'\mathbf{M} = \mathbf{0}$, $\mathbf{U}_p'\mathbf{L} = \mathbf{I}$.

Alternatively, the solution can be written as follows:

$$(10) \quad \begin{aligned}\mathbf{v} &= \mathbf{w}_p + \mathbf{L}^*\mathbf{U}_f'\mathbf{w}_f \\ &\quad + \mathbf{R}\mathbf{Z}_p\boldsymbol{\delta}\boldsymbol{\delta}'[\mathbf{I} + \mathbf{Z}_p'\mathbf{R}\mathbf{Z}_p\boldsymbol{\delta}\boldsymbol{\delta}']^{-1}[\mathbf{Z}_f' - \mathbf{Z}_p\mathbf{D}_p^{-1}\mathbf{U}_p(\mathbf{U}_p'\mathbf{D}_p^{-1}\mathbf{U}_p)^{-1}\mathbf{U}_f]\mathbf{w}_f,\end{aligned}$$

where $\mathbf{L}^* = \mathbf{D}_p^{-1}\mathbf{U}_p(\mathbf{U}_p'\mathbf{D}_p^{-1}\mathbf{U}_p)^{-1}$, $\mathbf{R} = \mathbf{D}_p^{-1} - \mathbf{D}_p^{-1}\mathbf{U}_p(\mathbf{U}_p'\mathbf{D}_p^{-1}\mathbf{U}_p)^{-1}\mathbf{U}_p'\mathbf{D}_p^{-1}$, so that $\mathbf{U}_p'\mathbf{L}^* = \mathbf{I}, \mathbf{U}_p'\mathbf{R} = \mathbf{0}$. The proof of the equivalence is direct.

It should be noticed that the DAF arises in the case $\mathbf{D} = \mathbf{K}^{-1}$ and $\mathbf{U} = \mathbf{X}$, so that the bias term is zero. When $\mathbf{D} = \sigma^2\mathbf{I}$ and $\boldsymbol{\delta}'(\mathbf{Z}_p'\mathbf{v} - \mathbf{Z}'\mathbf{w}) = 0$ (no bias term), an alternative equivalent derivation of the asymmetric filter approximating the two-sided local polynomial filter is based on the constrained minimization of the integrated squared modulus of the difference between the transfer function of the symmetric filter, denoted $\mathrm{w}(e^{-\imath\omega}) = \sum_{j=-h}^{h} \mathrm{w}_j e^{-\imath\omega j}$, where $\imath$ is the imaginary unit, and that of the asymmetric filter, denoted $\mathrm{v}(e^{-\imath\omega}) = \sum_{j=-q}^{h} \mathrm{v}_j e^{-\imath\omega j}$.

In particular, the asymmetric filter weights solve the following problem:

$$\min_{\mathrm{v}}\left\{\int_{-\pi}^{\pi} |\mathrm{w}(e^{-\imath\omega}) - \mathrm{v}(e^{-\imath\omega})|^2\, d\omega + \sum_{k=0}^{r} \lambda_k \left(\sum_{j=-q}^{h} j^k \mathrm{v}_j - \sum_{j=-h}^{h} j^k \mathrm{w}_j\right)\right\},$$

where $r$ can be equal to $1, 2, \ldots, d$ and $\lambda_k$ is a Lagrange multiplier. This approach has been applied for the construction of a filter approximating an ideal low-pass filter; see Percival and Walden (1993), Section 5.8 and Baxter and King (1999) for an application to the measurement of the business cycle.

4.1. *Musgrave asymmetric filters.* Musgrave's asymmetric filters [Musgrave (1964), Doherty (2001), Quenneville, Ladiray and Lefrancois (2003)] are obtained in the particular case when the original two-sided symmetric filter is the Henderson filter and $\mathbf{U} = \mathbf{i}$, $\mathbf{Z} = [-h, -h+1, \ldots, h]'$, $\boldsymbol{\delta} = \delta_1, \mathbf{D} = \sigma^2 \mathbf{I}$, that is, when $\mathbf{U}$ and $\mathbf{Z}$ are respectively the first and the second column of the design matrix $\mathbf{X}$.

It is nevertheless convenient for comparison purposes to reset the time origin and derive the filter under the equivalent design

$$\mathbf{U} = \mathbf{i}, \qquad \mathbf{Z} = [1, 2, \ldots, 2h+1]', \qquad \boldsymbol{\delta} = \delta_1, \qquad \mathbf{D} = \sigma^2 \mathbf{I}.$$

It is assumed that a linear process $y_t = \gamma_0 + \delta_1 t + \varepsilon_t$, $t = 1, \ldots, n$, $\mathrm{E}(\varepsilon_t) = 0$, $\mathrm{Var}(\varepsilon_t) = \sigma^2$, generates the observations, and that the asymmetric filter has to preserve a constant signal, that is, $\sum \mathrm{v}_i = 1$.

Then, if $M$ denotes the number of elements of $\mathbf{Z}_p$, $h < M < 2h+1$, and we let $H = 2h + 1$,

$$\mathbf{L}^* \mathbf{U}'_f \mathbf{w}_f = \frac{1}{M} \sum_{j=M+1}^{H} \mathrm{w}_j,$$

$$\mathbf{I} + \mathbf{Z}'_p \mathbf{R} \mathbf{Z}_p \boldsymbol{\delta} \boldsymbol{\delta}' = 1 + \frac{\delta_1^2}{\sigma^2} \frac{M(M+1)(M-1)}{12},$$

$$[\mathbf{Z}'_f - \mathbf{Z}_p \mathbf{D}_p^{-1} \mathbf{U}_p (\mathbf{U}'_p \mathbf{D}_p^{-1} \mathbf{U}_p)^{-1} \mathbf{U}_f] \mathbf{w}_f = \sum_{j=M+1}^{H} \left(j - \frac{M+1}{2}\right) \mathrm{w}_j,$$

$$\mathbf{R} \mathbf{Z}_p \boldsymbol{\delta} \boldsymbol{\delta}' = \frac{\delta_1^2}{\sigma^2} \left(i - \frac{M+1}{2}\right).$$

As a result, the usual expression for $\mathrm{v}_{[i]}$, the $i$th element of the vector $\mathbf{v}$, as presented in Doherty (2001), Findley et al. (1998) and Ladiray and Quenneville (2001), in terms of the elements of the vector $\mathbf{w} = \{\mathrm{w}_{[j]}, j = 1, \ldots, H\}$, is obtained:

(11)
$$\mathrm{v}_{[i]} = \mathrm{w}_{[i]} + \frac{1}{M} \sum_{j=M+1}^{H} \mathrm{w}_{[j]}$$
$$+ \frac{\delta_1^2}{\sigma^2} \left(i - \frac{M+1}{2}\right) \frac{\sum_{j=M+1}^{H} (j - (M+1)/2) \mathrm{w}_{[j]}}{1 + (\delta_1^2/\sigma^2) M(M+1)(M-1)/12},$$

for $i = 1, \ldots, M$.

The ratio $\frac{\delta_1^2}{\sigma^2}$ is related to $R = \bar{I}/\bar{C}$, as $\frac{\delta_1^2}{\sigma^2} = 4/(\pi R^2)$, where $\bar{I}$ is the expected absolute difference of the irregular component and $\bar{C}$ is the expected absolute difference of the trend component. Assuming $\varepsilon_t \sim \mathrm{NID}(0, \sigma^2)$, $|\varepsilon_t - \varepsilon_{t-1}|$ is half normal with expected value $4\sigma/\pi$ and $\bar{C} = \delta_1$, if the underlying signal is a linear trend with slope $\delta_1$.

The limit of (11) as $(\sigma^2/\delta_1^2) \to 0$ (or, equivalently, $\delta_1^2/\sigma^2 \to \infty$) is

$$\text{v}_{[i]} = \text{w}_{[i]} + \frac{1}{M} \sum_{j=M+1}^{N} \text{w}_{[j]}$$

(12)
$$+ \left( i - \frac{M+1}{2} \right) \frac{\sum_{j=M+1}^{N} (j - (M+1)/2) \text{w}_{[j]}}{M(M+1)(M-1)/12},$$

for $i = 1, \ldots, M$. This expression corresponds to what we would obtain if the unavailable future observations were replaced by linear extrapolations formed from the available data. See also [Doherty (2001), Section 6].

4.2. *The properties of the approximate filters.* The approximate filters that minimize (8) raise a controversial point. The symmetric filter was derived from the assumption that the series behaves locally according to a polynomial of degree $d$. We seek to approximate this filter by changing our assumption about how $y_t$ is generated, postulating that it has been possibly generated by a lower order polynomial or that the asymmetric filter is only capable of reproducing a polynomial of lower degree. In one way or another we are denying the conditions under which the original smoothing filter was derived. However, it is clear from our previous discussion that the original motivation for introducing a new class of approximating filters was the fact that the direct real time filter delivers very volatile estimates; hence, we had to move away from the direct strategy of fitting the maintained polynomial to the available observations. Second, it is not implausible to assume that the behavior of the signal at the extremes is different from that in the interior of the sampling design. An analogy can be drawn with cubic smoothing splines: the so-called natural boundary condition is such that the spline is a local cubic function of time inside the boundary and is linear outside. This is beneficial to the reliability of the real time estimates and of the forecasts. The strategy that is adopted in the approximation is very similar since it effectively amounts to reducing the order of the fitting polynomial.

The merits of the class of filters (10), relative to the DAF, lie in the bias-variance trade-off. In particular, the bias can be sacrificed for improving the variance properties of the corresponding asymmetric filter. If $\mathbf{U} = \mathbf{X}$, that is, it is asked of the filter to be capable of reproducing a $d$th order polynomial, which is also the process generating the observations, the approximate filter minimizing $\varphi(\mathbf{v})$ will not differ from $\mathbf{w}_a$ (in the light of the result in Section 3.2 they will coincide if $\mathbf{D} = \mathbf{K}^{-1}$); as a consequence, the real time filter will be strongly localized, and it will suffer from the same limitations as the DAF discussed in Section 3.1, namely, its estimates will be characterized by high variance.

When $\mathbf{U}$ is a subset of the columns of $\mathbf{X}$, spanning a polynomial of degree $d^* < d$, then we require that the filter is capable of reproducing a polynomial of degree $d^*$; if the observations are generated by a polynomial of degree greater than $d^*$, a bias will arise, which depends on the value of $\boldsymbol{\delta}$. However, the weights of the approximating filter will be more evenly distributed and the variance will be reduced. Thus, the overall mean square revision error may eventually be reduced if the actual signal is weakly evolutive.

Figure 4 plots the gain and the phase function of the real time filter when $h = 6$, $\mathbf{w}$ is the two-sided Henderson filter, and $\mathbf{U} = \mathbf{i}$ (the asymmetric weights have to satisfy the constraint $\sum v_i = 1$), $\mathbf{Z} = [-h, -h+1, \ldots, h-1, h]'$, $\boldsymbol{\delta} = \delta_1, \mathbf{D} = \sigma^2 \mathbf{I}$. As the filter depends on the slope of the underlying signal through the ratio $\delta_1^2/\sigma^2$, we plot two limiting cases arising when the slope is negligible and when it is the dominating feature. The intermediate case is the well-known Musgrave surrogate real time filter for the Henderson with 13 terms, which rises when $\delta_1^2/\sigma^2 = 4/(3.5^2\pi)$, with $R = 3.5$ being the value selected for the Henderson filter with 13 terms. The filter weights are displayed in the bottom right panel of the same figure.

When the slope is negligible, $\delta_1^2/\sigma^2 = 0$ (or, equivalently, $R \to \infty$, which arises either when the signal is constant and is devoid of the linear term or the signal is buried in a heap of noise), the optimal approximation to the Henderson two-sided filter features weights that are less dispersed and the gain decreases from 1 almost monotonically, as it ought to be expected. The individual weights of the real time filter, $v_0, \ldots, v_h$, are plotted against the value of $\delta_1^2/\sigma^2$ in the bottom left panel. As the linear signal is stronger, the dispersion of the weights increases and the gain becomes higher at each individual frequency, getting greater than one at the low frequencies.

The class of filters (10) accommodates the case when the two-sided symmetric filter is the Henderson filter and an approximation is sought such that for a locally quadratic underlying signal, $y_{t+j} = \gamma_0 + \gamma_1 j + \delta_2 j^2 + \varepsilon_{t+j}$, $\varepsilon_{t+j} \sim \text{IID}(0, \sigma^2)$, and requiring that the approximating filter preserves a linear signal, which is achieved by imposing the constraints $\sum v_j = 1, \sum_{j=-q}^{h} v_j j = \sum_{j=-h}^{h} w_j j$. In (8) we set

$$\mathbf{U}' = \begin{bmatrix} 1 & 1 & \cdots & 1 & \cdots & 1 & 1 \\ -h & -h+1 & \cdots & 0 & \cdots & h-1 & h \end{bmatrix},$$

and

$$\mathbf{Z} = [(-h)^2, (-h+1)^2, \ldots, 1, 0, 1, \ldots, h^2]', \qquad \boldsymbol{\delta} = \delta_2, \qquad \mathbf{D} = \sigma^2 \mathbf{I}.$$

Hence, $\mathbf{U}$ consists of the first two columns of $\mathbf{X}$ and $\mathbf{Z}$ is the third column, and the filter weights depend on the curvature of the underlying signal via $\delta_2^2/\sigma^2$. The filter $\mathbf{v}$ will be referred to as the quadratic trend–linear fit (QL) approximation to the Henderson filter.

The first row of Figure 5 considers the real time QL filter and plots the gain, the phase and the individual filter weights for different values of $\delta_2^2/\sigma^2$, which expresses the relative importance of the curvature of the signal. The rationale for this particular type of asymmetric filter is early detection of turning points, which are a quadratic feature of the signal; setting $\delta_2/\sigma^2$ to a high value, the optimal filter would weight more the current observation and detect a turning point more rapidly. Essentially, with respect to the Musgrave type of filters, the bias is reduced at the expense of the variance. It should also be noticed that the optimal filter for $\delta_2^2/\sigma^2 = 0$ is coincident with the optimal filter derived under a linear trend signal and using the constraint $\sum_i v_i = 0$ with $\delta_1^2/\sigma^2 \to \infty$; compare Figure 4.

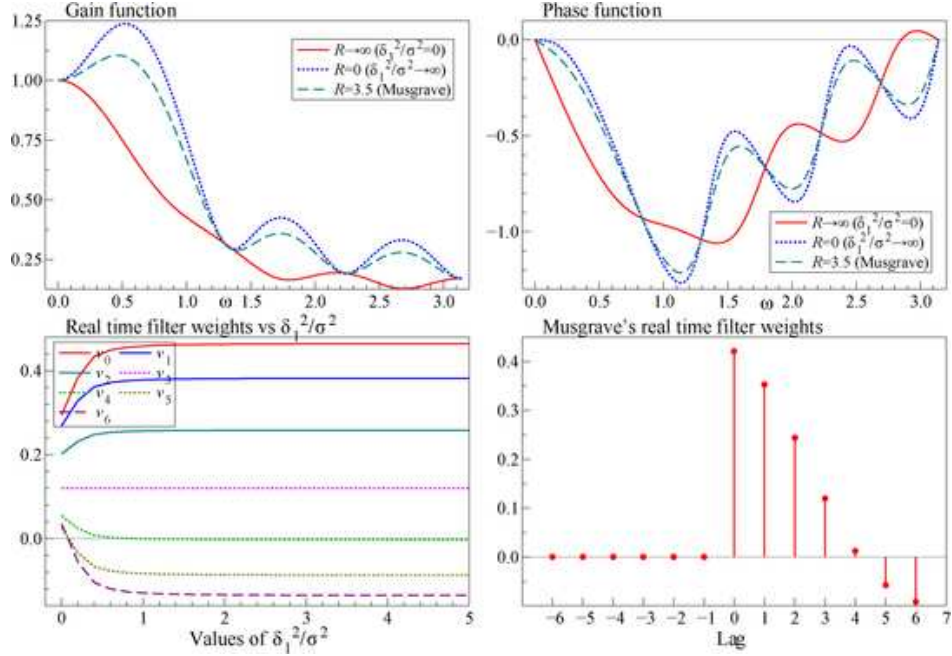

Fig. 4. *Gain, phase and weights for the real time filter minimizing the revision mean square error subject to $\sum v_i = 1$, when the observations are generated by a linear trend.*

Finally, the bottom panels of Figure 5 display the gain, phase and filter weights of the real time filter approximating the Henderson filter when the series is generated by a cubic polynomial, $y_{t+j} = \gamma_0 + \gamma_1 j + \gamma_2 j^2 + \delta_3 j^3 + \varepsilon_{t+j}$, and the weights have to satisfy the quadratic reproduction constraints $\sum v_j = 1$, $\sum_{j=-q}^{h} v_j j = \sum_{j=-h}^{h} w_j j$ and $\sum v_j j^2 = \sum_{j=-h}^{h} w_j j^2$. These filters will be referred to the cubic trend—quadratic fit (CQ) asymmetric filters. In this case $\mathbf{U}$ is a matrix formed from the first three columns of the $\mathbf{X}$ matrix,

$$\mathbf{U}' = \begin{bmatrix} 1 & 1 & \cdots & 1 & \cdots & 1 & 1 \\ -h & -h+1 & \cdots & 0 & \cdots & h+1 & h \\ h^2 & (-h+1)^2 & \cdots & 0 & \cdots & (h+1)^2 & h^2 \end{bmatrix},$$

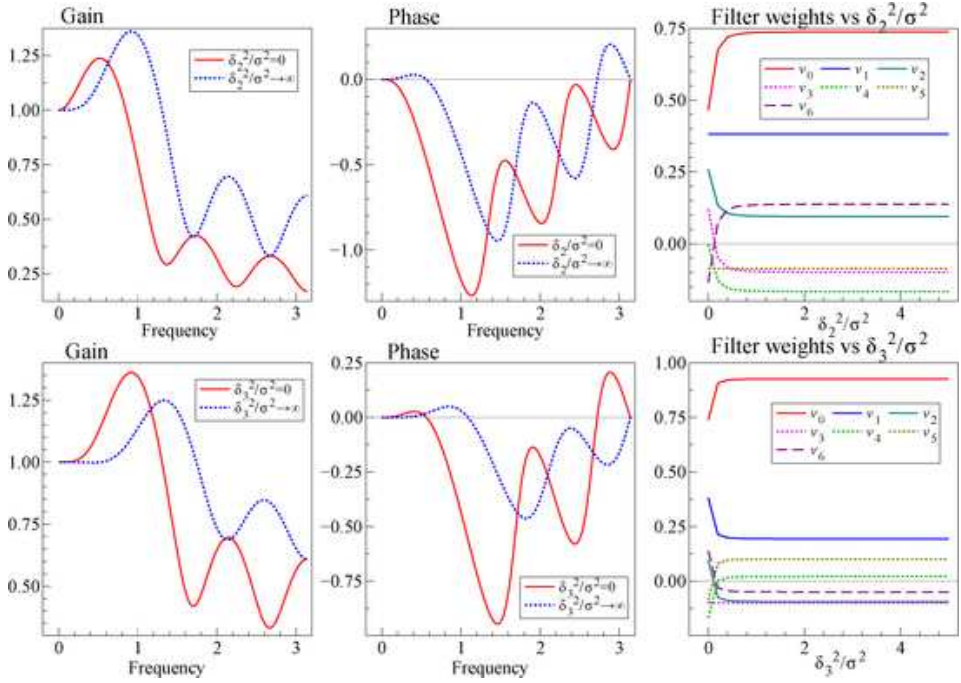

Fig. 5. *Gain, phase and weights for the real time filter minimizing the revision mean square error: QL and CQ filters.*

and

$$\mathbf{Z} = [(-h)^3, (-h+1)^3, \ldots, 1, 0, 1, \ldots, h^3]', \qquad \boldsymbol{\delta} = \delta_3, \qquad \mathbf{D} = \sigma^2 \mathbf{I}.$$

In this case the optimal filter depends on the parameter $\delta_3^2/\sigma^2$, which is a measure of relative inflexion. Again, the optimal filter for $\delta_3^2/\sigma^2 = 0$ is the same as the QL filter arising for $\delta_2^2/\sigma^2 \to \infty$; compare the top panels of Figure 5. As $\delta_3^2/\sigma^2 \to \infty$, the filter is the same as the direct asymmetric filter of Section 3.1.

**5. Illustrations.** In this section we provide illustrations concerning the use of the general expression (9) for the design of real time filters suitable for a particular time series. The reference two-sided filter is the Henderson filter and we estimate the bandwidth $h$ by cross-validation.

Let $\hat{m}_{t\backslash t}$ denote the two-sided estimate of the signal at time $t$ which does not use $y_t$. The latter can be expressed in terms of the Henderson estimate of the trend using the central filter:

$$\hat{m}_{t\backslash t} = \mathbf{e}_1'(\mathbf{X}'\mathbf{K}\mathbf{X} - \kappa_0 \mathbf{e}_1 \mathbf{e}_1')^{-1}(\mathbf{X}'\mathbf{K}\mathbf{y} - \kappa_0 y_t \mathbf{e}_1)$$

$$= \mathbf{e}_1' \left[ (\mathbf{X}'\mathbf{K}\mathbf{X})^{-1} + \frac{\kappa_0}{1 - \kappa_0 \mathbf{e}_1'(\mathbf{X}'\mathbf{K}\mathbf{X})^{-1}\mathbf{e}_1} (\mathbf{X}'\mathbf{K}\mathbf{X})^{-1} \mathbf{e}_1 \mathbf{e}_1' (\mathbf{X}'\mathbf{K}\mathbf{X})^{-1} \right]$$

$$\times (\mathbf{X}'\mathbf{K}\mathbf{y} - \kappa_0 y_t \mathbf{e}_1)$$
$$= \frac{1}{1-\mathrm{w}_0}\mathbf{e}_1'(\mathbf{X}'\mathbf{K}\mathbf{X})^{-1}(\mathbf{X}'\mathbf{K}\mathbf{y} - \kappa_0 y_t \mathbf{e}_1)$$
$$= \frac{1}{1-\mathrm{w}_0}\hat{m}_t - \frac{\mathrm{w}_0}{1-\mathrm{w}_0}y_t.$$

The leave-one-out, or deletion, residual can be expressed in terms of the trend estimate using all the observations:

$$y_t - \hat{m}_{t\setminus t} = \frac{1}{1-\mathrm{w}_0}(y_t - \hat{m}_t).$$

The cross-validation score is the sum of the squared deletion residuals:

$$CV = \sum_{t=h+1}^{n-h} (y_t - \hat{m}_{t\setminus t})^2 = \sum_{t=h+1}^{n-h} \frac{(y_t - \hat{m}_t)^2}{(1-\mathrm{w}_0)^2}.$$

Conditional on the value of $h$ we consider three classes of filters:

ASYMMETRIC LC. The asymmetric LC (Linear-Constant) real time filter arises as the best approximation to the two-sided Henderson filter, assuming that $y_t$ is linear and imposing the constraint that the weights sum to 1. Hence, $\mathbf{U} = \mathbf{i}$, the unit vector, and the asymmetric filter depends on $\delta_1$; see expression (11).

ASYMMETRIC QL. The asymmetric QL (Quadratic-Linear) real time filter arises as the best approximation to the two-sided Henderson filter, assuming that $y_t$ is quadratic and imposing the constraint that the estimates are capable of reproducing a first degree polynomial (see Section 4.2).

ASYMMETRIC CQ. The asymmetric CQ (Cubic-Quadratic) real time filter arises as the best approximation to the two-sided Henderson filter, assuming that $y_t$ is a cubic function of time and imposing the constraint that the estimates are capable of reproducing a second degree polynomial (see Section 4.2).

The three asymmetric filters depend on a single parameter, $\delta_i^2/\sigma^2$, $i = 1, 2, 3$. For each we compute the value that minimizes the mean square revision error (MSRE), that is, the value for which $\sum_{t=h+1}^{n-h}(\hat{m}_t - \hat{m}_{t|t})^2/(n-2h-1)$ is a minimum.

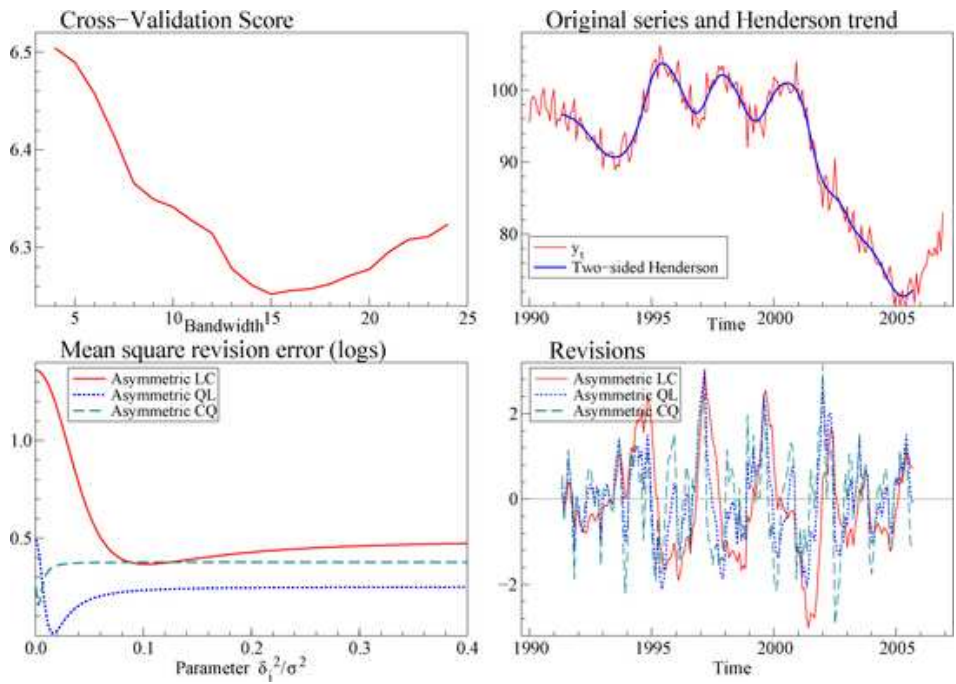

FIG. 6. *Selection of a real time filter: Index of Industrial Production, Italy, branch DL. Source: Istat.*

5.1. *Italian index of industrial production.* Our first illustration deals with the Italian index of industrial production for the branch DL (Manufacture of electrical and optical equipment, Nace Rev. 1 classification). The series is produced by ISTAT, the Italian National Statistical Office, and made available on the website www.istat.it. The index shows the evolution of gross production in volume terms and represents a key short term indicator, due also to its timeliness, being made available with a delay of 43 days after the end of the reference month. The data are collected monthly through a survey of establishments with at least 20 employees that make up at least 70% of total production. The volume of production in month $t$ is compared to the average production of the base year (2000 for the current release). This dataset is available as supplementary material [Proietti and Luati (2008)], along with the other time series used for our illustrations.

The value of the bandwidth selected by cross-validation is $h = 15$; the two-sided estimates of the trend are displayed in the right top panel of Figure 6. We next look for the best approximation to the Henderson filter within the three particular classes. For this purpose we estimate the values of the parameters $\delta_i^2/\sigma^2$, $i = 1, 2, 3$, using a grid search. The results are presented in the bottom left panel of Figure 6. The minimizers of the MSRE are $\widehat{(\delta_1^2/\sigma^2)} = 0.103$, $\widehat{(\delta_2^2/\sigma^2)} = 0.016$ and $\widehat{(\delta_3^2/\sigma^2)} = 0.003$, respectively, for the LC, QL and CQ classes. As illustrated by Figure 6, the best approximation to the original Henderson filter is provided by the QL filter with $\widehat{(\delta_2^2/\sigma^2)} = 0.016$. We need the real time filter to be capable of reproducing a linear signal and to react somewhat, although not in full, to the curvature of the underlying trend.

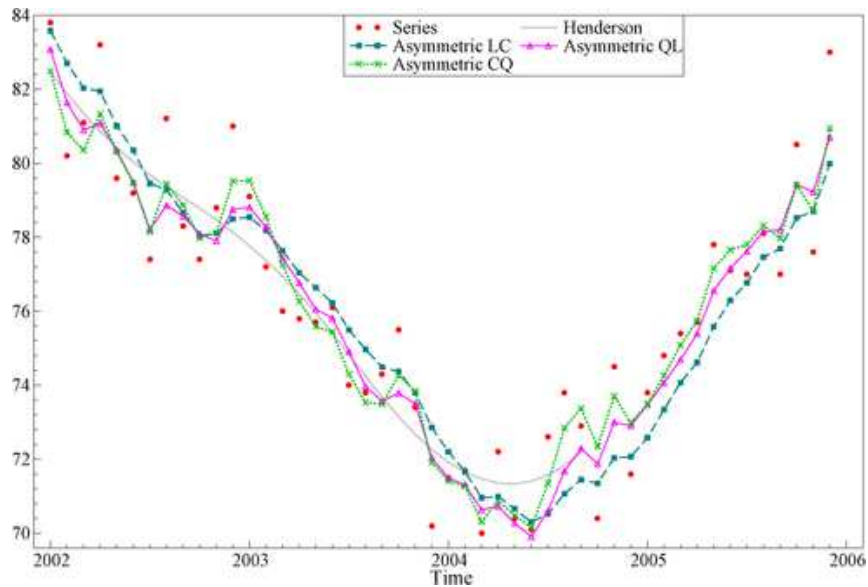

Fig. 7. *Index of Industrial Production, Manufacture of electrical and optical equipment, Italy. Comparison of the real time estimates arising from three approximating filters and final estimates of the trend component.*

Figure 7 compares the real time estimates of the trend for the period January 2002–December 2006, $\hat{m}_{t|t}$, arising from the best LC, QL and CQ approximations. It it is clear that the LC filter is biased when the slope is substantial: the bias is positive in a recessionary period and negative in expansion. This is so since the filter can only preserve a constant, but will distort a local linear trend. The optimal QL approximation provides the best approximation since the real time estimates are closer to the final Henderson estimates. The CQ approximation tracks the data quite well, but the corresponding estimates are affected by higher variance, compared with the QL estimates. Similar considerations apply to the DAF estimates, not reported for brevity. For the class of economic time series that are usually considered, such as industrial production, the evidence definitively points out that the direct asymmetric filter produces the most inefficient estimates, due to the very high variance inflation.

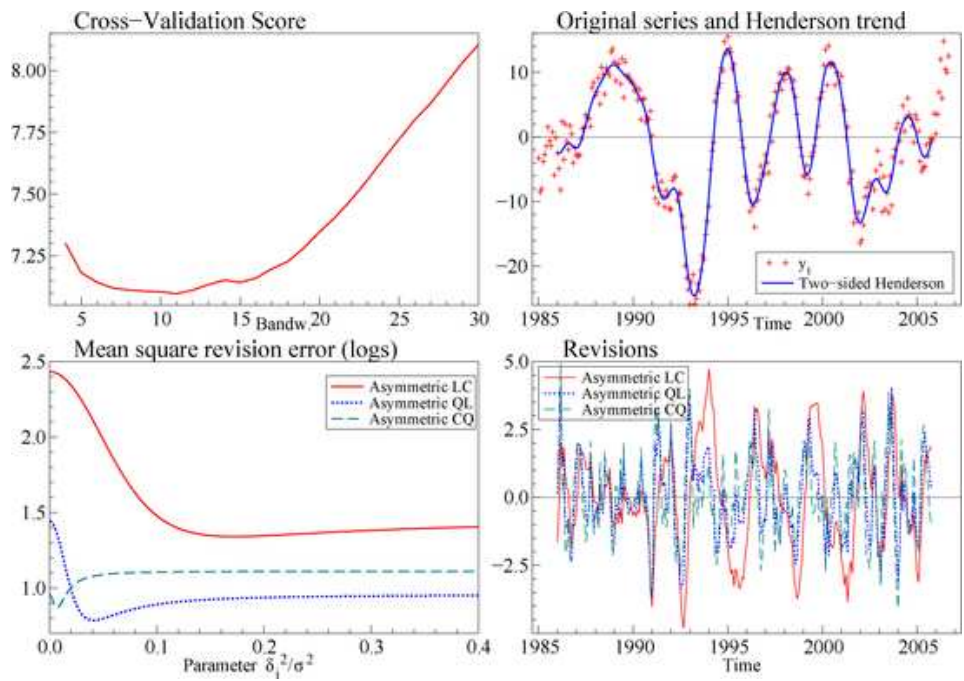

Fig. 8. *Selection of a real time filter: Assessment of order-book levels, Euro Area. Source: European Commission.*

5.2. *Assessment of order-book levels.* Our second illustration deals with the monthly assessment of order-book levels for the 13 countries constituting the Euro area. The series is produced by the European Commission, Directorate General for Economic and Financial Affairs, which conducts a monthly survey of the industrial sector of the economies in the European Union. The survey is largely qualitative and is administered to a purposive sample of about 23,000 representative firms. The main questions refer to an assessment of recent trends in production, of the current levels of order books and stocks, as well as expectations about production, selling prices and employment. The survey question from which our series originates is whether over the past three months the firm's orders have increased, remained unchanged or decreased. Answers obtained from the surveys are aggregated in the form of balances, which are constructed as the difference between the percentages of respondents giving positive and negative replies. See European Commission (2007) for more details.

The series is made available on the website http://ec.europa.eu/economy_finance/db_indicators and is plotted in the second panel of Figure 8. It provides an interesting case study, since its dynamic behavior is highly cyclical. The sample period considered is January 1985–September 2006.

As it can be seen from the first panel of Figure 8, the value of the bandwidth parameter suggested by cross-validation is $h = 11$. The two-sided estimates of the trend resulting from the Henderson filter corresponding to the selected $h$ value are plotted in the right top panel. The mean square revision error for the three filters is plotted in the bottom left panel against the value of $\delta_i^2/\sigma^2$, $i = 1, 2, 3$. The minimizers of the MSRE are $\widehat{(\delta_1^2/\sigma^2)} = 0.173$, $\widehat{(\delta_2^2/\sigma^2)} = 0.041$ and $\widehat{(\delta_3^2/\sigma^2)} = 0.007$, respectively. Overall, the MSRE is minimized by the QL filter, which again is our preferred real time filter. The bottom right panel displays the revisions $\hat{m}_{t|t} - \hat{m}_t$ for the optimal filters belonging to each subclass, showing that the latter are particularly large for the LC filter.

Figure 9 compares the real time estimates of the trend component for the period January 2002–September 2006, arising from the best LC, QL and CQ approximations. The series is characterized in this period by the presence of several turning points and by the rapid alternation of different business cycle phases. The plot illustrates that, due to the asymmetry of the real time filters, all the real time trend estimates suffer from a displacement of the turning points along the time axis, also known as a phase shift; the best performance is, however, provided by the QL approximation. The LC filter can depart quite substantially both from the final trend estimates and from the actual series values, during the phases of steep recovery after a lower turning point; on the other hand, the asymmetric CQ estimates are too responsive to the observations and suffer from excess volatility.

5.3. *Housing starts.* We conclude with a more detailed treatment of the series concerning the number of new housing units started in the US, considered in Section 2 and depicted in Figure 1. The value of the bandwidth estimated by cross-validation is $h = 10$ (the cross-validation score is presented in the top left panel of Figure 10), and, thus, the Henderson estimates of the trend, displayed in the top right panel, are based on 21 consecutive observations. For the estimation (at the beginning and) at the end of the sample period, the best asymmetric approximation is provided by the QL filter with $\widehat{(\delta_2^2/\sigma^2)} = 0.029$.

Figure 11 compares the three real time estimates of the trend component. Those yielded by the QL asymmetric filter provide the best compromise between flexibility and smoothness: they are indeed more flexible than the LC estimates, which is particularly advantageous during the last steep recession initiated in 2006, but far less volatile than the CQ estimates, which, on the contrary, are too sensitive to the influence exerted by the current observations.

In conclusion, the evidence presented in this section illustrates that the proposal of designing asymmetric filters in a more general and flexible way helps estimating the underlying signal with greater accuracy.

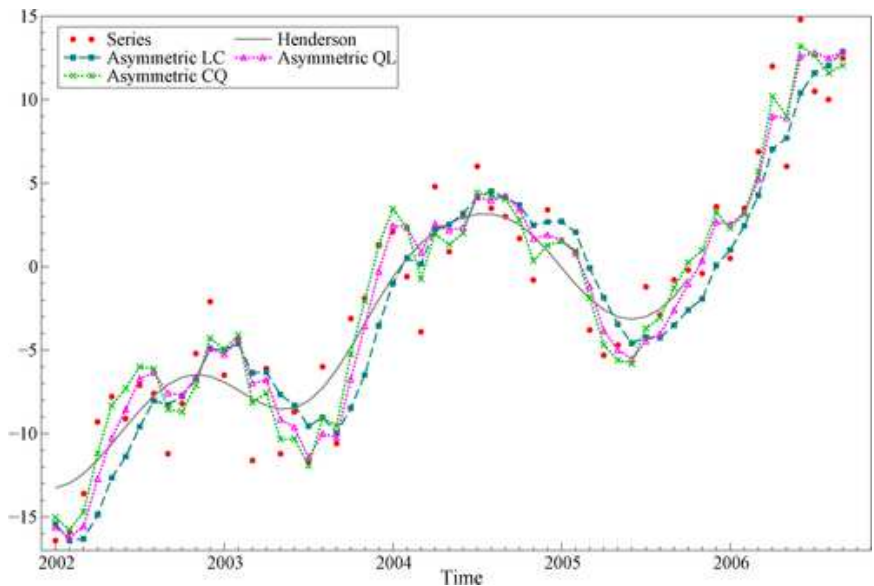

Fig. 9. *Assessment of order-book levels, Euro Area. Comparison of the real time estimates arising from three approximating filters and final estimates of the trend component.*

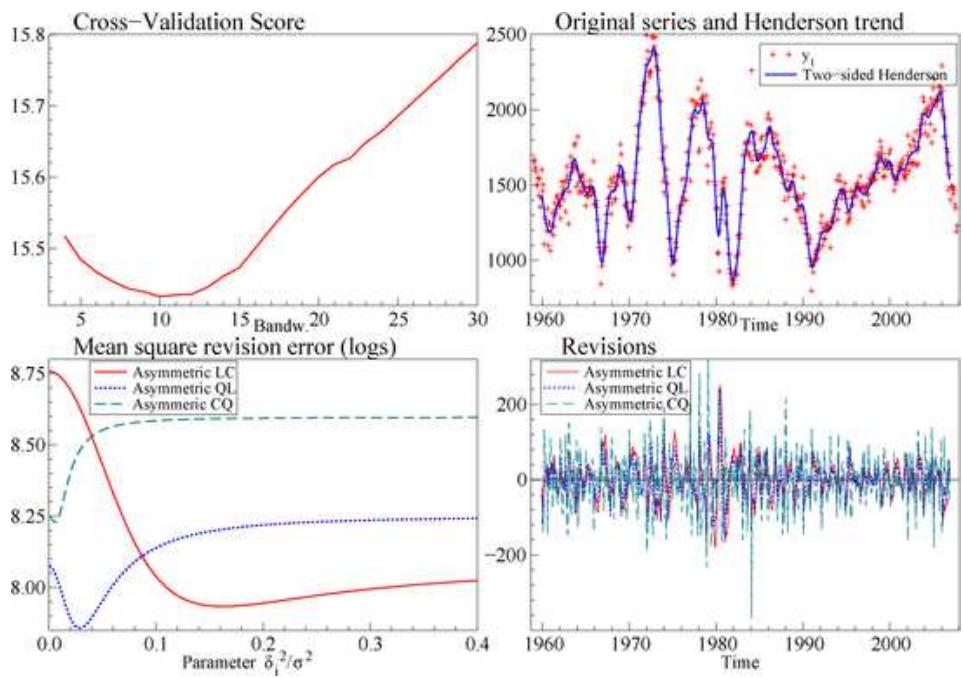

Fig. 10. *Selection of a real time filter: US Housing Starts. Source: European Commission.*

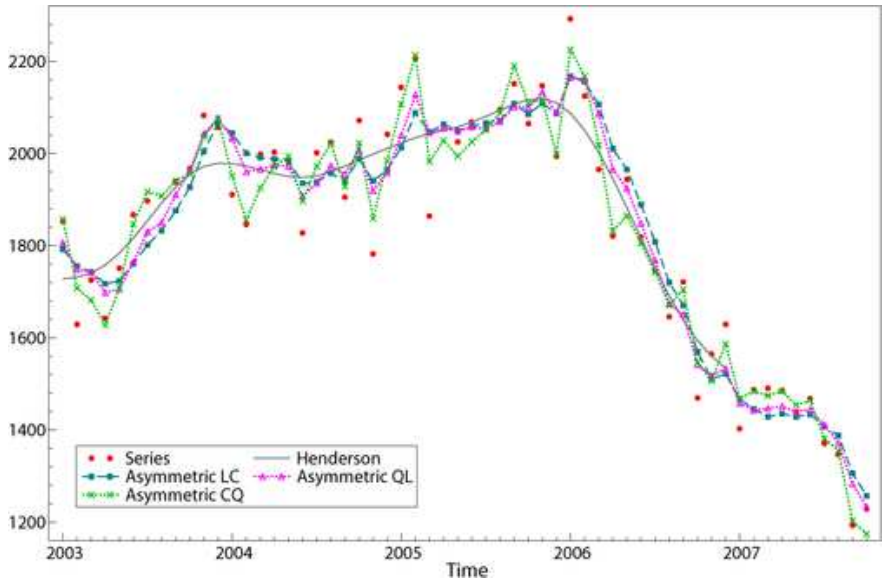

Fig. 11. *US Housing Starts. Comparison of the real time estimates arising from three approximating filters and final estimates of the trend component.*

**6. Conclusions.** The paper has considered the problem of estimating the trend of a time series in real time by means of local polynomial filters; we showed that automatic adaptation at the boundary fails due to the high volatility of the estimates. We thus evaluated the strategy of approximating a given symmetric local polynomial filter by minimizing the mean square revision error subject to different order polynomial reproducing constraints and by making certain assumptions concerning the nature of the underlying signals. Restricting our attention to three families of real time filters that depend on certain key features of the unknown signal, such as its slope and curvature, we proposed to estimate these key features from the available data, rather than taking a fixed filter.

Our empirical illustrations concerned the minimum mean square revision error approximation of the Henderson filter, a very popular local cubic smoother. They enable us to conclude that we can improve upon the well-known Musgrave asymmetric filters, which for the series considered suffer from very large revisions, especially in steep recessions and recoveries and around turning points. This evidence arises as a consequence of the fact that the filter is not designed to deal with signals characterized by strong slope and curvature.

We also considered the strategy of building either direct or minimum revision mean square approximations using the same fixed number of observations (nearest neighbor bandwidth), rather than a fixed bandwidth.

The nearest neighbor bandwidth, proposed by Cleveland (1979), has certain advantages over the fixed bandwidth, and, in particular, when the observations become sparse. In fact, it can be shown that the minimum mean square asymmetric approximation has better theoretical properties than the fixed bandwidth counterpart, but its effectiveness in ameliorating the approximation to the Henderson filter was not proven by our empirical applications.

## APPENDIX

When $d < h$ (or $d < 2h$ in the nearest neighbor case, where $q = h + 1$ is fixed) and $h + q + 1$ is the asymmetric filter length varying with $q = 0, \ldots, h - 1$, the generalized Binet–Cauchy formula can be used to determine how the leverage varies with $d$ or $h$:

$$\frac{\det(\mathbf{M}_{1,1})}{\det(\mathbf{X}'_p \mathbf{K}_p \mathbf{X}_p)} = \frac{\sum_{\pi_j}^{\binom{h+q}{d}} \det(\mathbf{X}'_{h+1,1 \cdot \pi_j}) \det(\mathbf{K}_{h+1,h+1_{\pi_j}}) \det(\mathbf{X}_{h+1,1_{\pi_j} \cdot})}{\sum_{\pi_j}^{\binom{h+q+1}{d+1}} \det(\mathbf{X}'_{\cdot \pi_j}) \det(\mathbf{K}_{\pi_j}) \det(\mathbf{X}_{\pi_j \cdot})},$$

(13)

where $\mathbf{X}'_{\cdot \pi_j}$ denotes a square submbatrix of $\mathbf{X}'_p$ obtained taking all its rows and $d + 1$ columns chosen on the set $\pi_j$ of the $h + q + 1$ columns of $\mathbf{X}'_p$ and the summation $\sum_{\pi_j}^{\binom{h+q+1}{d+1}}$ is extended to the $\binom{h+q+1}{d+1}$ subsets of $1, \ldots, h+q+1$ with $d+1$ elements; $\mathbf{K}_{\pi_j}$ is the square submatrix of $\mathbf{K}_p$ whose $d+1$ columns (and rows) correspond to those chosen for $\mathbf{X}'_{\cdot \pi_j}$. The matrices $\mathbf{X}'_{h+1,1 \cdot \pi_j}$ and $\mathbf{K}_{h+1,h+1_{\pi_j}}$ are of dimension $d \times d$.

It is immediate to verify that, for $q = d = 0$, the ratio is equal to $(\sum_{j=0}^{h} \kappa_j)^{-1} = S_{00}^{-1}$. On the other extreme, for $q = 0, d = h$, we find the classical Binet–Cauchy formula for square matrices giving ratio equal to $\kappa_0^{-1}$. In the interior $0 < d < h$ the ratio (13) becomes

$$\frac{\det(\mathbf{M}_{1,1})}{\det(\mathbf{X}'_p \mathbf{K}_p \mathbf{X}_p)} = \sum_{\pi_j}^{\binom{h+q}{d}} \det(\mathbf{X}'_{h+1,1 \cdot \pi_j})^2 \det(\mathbf{K}_{h+1,h+1_{\pi_j}})$$

$$\times \left[ \sum_{\pi_j}^{\binom{h+q}{d}} \det(\mathbf{X}'_{\cdot \pi_{j(h+1)}})^2 \det(\mathbf{K}_{\pi_{j(h+1)}}) \right.$$

$$\left. + \sum_{\pi_j}^{\binom{h+q+1}{d+1} - \binom{h+q}{d}} \det(\mathbf{X}'_{\cdot \pi_{j/(h+1)}})^2 \det(\mathbf{K}_{\pi_{j/(h+1)}}) \right]^{-1},$$

where $\pi_{j(h+1)}$ indicates that only the submatrices of $\mathbf{X}'_p$ or $\mathbf{K}_p$ with the column $h+1$ are considered, while $\pi_{j/(h+1)}$ indicates that the column $h+1$ is not included. Since $\det(\mathbf{X}'_{\cdot\pi_{j(h+1)}}) = \pm \det(\mathbf{X}'_{h+1,1.\pi_j})$, for even or odd values of $d+2$, respectively, and $\det(\mathbf{K}_{\pi_{j(h+1)}}) = \kappa_0 \det(\mathbf{K}_{h+1,h+1_{\pi_j}})$, then $\sum_{\pi_j}^{\binom{h+q}{d}} \det(\mathbf{X}'_{\cdot\pi_{j(h+1)}})^2 \det(\mathbf{K}_{\pi_{j(h+1)}}) = \kappa_0 \sum_{\pi_j}^{\binom{h+q}{d}} \det(\mathbf{X}'_{h+1,1.\pi_j})^2 \det(\mathbf{K}_{h+1,h+1_{\pi_j}})$, so that

$$w_0^a = \sum_{\pi_j}^{\binom{h+q}{d}} \det(\mathbf{X}'_{\cdot\pi_{j(h+1)}})^2 \det(\mathbf{K}_{\pi_{j(h+1)}})$$

$$\times \left[ \sum_{\pi_j}^{\binom{h+q}{d}} \det(\mathbf{X}'_{\cdot\pi_{j(h+1)}})^2 \det(\mathbf{K}_{\pi_{j(h+1)}}) \right.$$

$$\left. + \sum_{\pi_j}^{\binom{h+q+1}{d+1}-\binom{h+q}{d}} \det(\mathbf{X}'_{\cdot\pi_{j/(h+1)}})^2 \det(\mathbf{K}_{\pi_{j/(h+1)}}) \right]^{-1}.$$

The above expression enables to write (13) and, consequently, $w_0^a$ as a function of determinants which are positive and refer to matrices having the same dimensions, $d+1 \times d+1$. Hence, it follows that $w_0^a$ increases (decreases) if the value $\sum_{\pi_j}^{\binom{h+q+1}{d+1}-\binom{h+q}{d}} \det(\mathbf{X}'_{\cdot\pi_{j/(h+1)}})^2 \det(\mathbf{K}_{\pi_{j/(h+1)}})$ decreases (increases). The latter is made of $\binom{h+q}{d+1}$ positive terms, so that it is sufficient to evaluate how this number of terms varies by varying $d, q, h$ to determine how $w_0^a$ varies accordingly. Given that $d < h$ and $q > 0$, then $h+q \geq d+1$ and, therefore:

(i) for fixed $d$, an increase in $q$ or $h$ implies an increase in $\binom{h+q}{d+1}$, that is, a decrease in $w_0^a$;

(ii) for fixed $h$, an increase in $d$ implies a decrease in $\binom{h+q}{d+1}$, that is, an increase in $w_0^a$.

**Acknowledgments.** This paper has been presented at the International Workshop on Computational and Financial Econometrics, Geneva, Switzerland, April 20–22, 2007, and at the 56th Session of the ISI 2007 Lisbon, August 22–29, 2007. We are especially grateful to Estela Bee Dagum, David Findley, Domenique Ladiray, Gianluigi Mazzi, Benoit Quenneville and an Associate Editor for their very competent comments. Financial support from MIUR (Ministero dell'Università e della Ricerca), Prin 2004, is gratefully acknowledged.

## SUPPLEMENTARY MATERIAL

**Datasets** (DOI: 10.1214/08-AOAS195SUPP; .zip). The supplementary material contains the time series used to illustrate the methods. The series are the following: assessment of order-book levels, housing starts: total: new privately owned housing units started, and index of industrial production, branch DL, manufacture of electrical and optical equipment

University of Rome "Tor Vergata"
S.E.F. e ME. Q.
via Columbia 2
00133 Roma
Italy
E-mail: tommaso.proietti@uniroma2.it

University of Bologna
Department of Statistics
via Belle Arti 41
40126 Bologna
Italy
E-mail: alessandra.luati@unibo.it